\documentstyle[aaspp4]{article}

\newcommand{\etal}{{\it et al. \,}}
\newcommand{\kms}{km~s$^{-1}$}
\def\T1{\ {$T_1$}\ }
\def\CT1{\ {$(C-T_1)$}\ }

\def\gtsim{\ {\raise-0.5ex\hbox{$\buildrel>\over\sim$}}\ }
\def\ltsim{\ {\raise-0.5ex\hbox{$\buildrel<\over\sim$}}\ }

\lefthead{Lee {\it et al.}}
\righthead{Globular Cluster System in NGC 4472}

\begin{document}

\title{Washington Photometry of the Globular Cluster System of NGC 4472.\\
 II. The Luminosity Function and Spatial Structure
}

\author{Myung Gyoon Lee$^1$, Eunhyeuk Kim}
\affil{Department of Astronomy, Seoul National University, Seoul 151-742, 
Korea \\
Electronic mail: mglee@astrog.snu.ac.kr, ekim@astro.snu.ac.kr}

\and

\author{Doug Geisler$^1$}
\affil{KPNO/NOAO, Tucson, AZ 85719, USA \\
Electronic mail: dgeisler@noao.edu}

\altaffiltext{1}{Visiting Astronomer, Kitt Peak National Observatory, National
Optical Astronomy Observatories, operated by the Association of 
Universities for Research in Astronomy, Inc., under contract with the
National Science Foundation.
}

\begin{abstract}

We present a comprehensive 
study of the luminosity function and spatial structure of the
globular cluster system of NGC 4472, the brightest galaxy in Virgo, 
based on deep wide field Washington $CT_1$ CCD images.
The globular cluster luminosity function shows a peak at $T_1=23.3\pm0.1$ mag,
about 1.5 mags brighter than our 50\% completeness limit.
Comparing this value with that of the Galactic globular clusters, we estimate 
the true distance modulus to NGC 4472 to be $(m-M)_0=31.2\pm 0.2$ 
(corresponding to a distance of $17.4\pm 1.6$ Mpc). 
With our large sample ($\approx 2000$) of bright globular clusters over a wide field, 
we make a definitive investigation
of the spatial structures of the metal-poor and metal-rich cluster
populations and find that they are systematically different:
(1) The metal-rich clusters are
more centrally concentrated than the metal-poor clusters;
and (2) The metal-rich clusters are elongated roughly
along the major axis of the parent galaxy,
while the metal-poor clusters are essentially spherically distributed.
In general, the metal-rich clusters closely
follow the underlying halo starlight of NGC 4472 
in terms of spatial structure and metallicity, 
while the metal-poor clusters do not.
The global value of the specific frequency of the globular clusters in NGC 4472
is estimated to be $S_N=4.7 \pm 0.6$.
The local specific frequency increases linearly
outward from the center of NGC 4472
until $\sim 5\arcmin.5$, beyond which
it levels off at $S_N\sim 8.5$ until the limit of our data at $7\arcmin$.
The specific frequency of both the metal-rich and metal-poor populations 
shows similar behavior.
However, $S_N$ of the metal-poor clusters is about a factor 2 larger than
that of the metal-rich clusters in the outer regions. 
Implications of these results for 
the origin of the globular clusters in NGC 4472 
are discussed.
These results are consistent with many of the predictions of both the model of 
episodic {\it in situ} formation plus tidal stripping of globular clusters 
given by Forbes \etal [1997, AJ,  113, 1652] and the Ashman \& Zepf [1992, 
ApJ, 384, 50]
merger formation model, but each of the models also has some problems.

\end{abstract}

\section{INTRODUCTION}

This is the second in a series of papers on Washington CCD photometry
of the globular cluster system (GCS) in NGC 4472. 
In Paper I (\cite{gei96})
we obtained  a deep color-magnitude diagram of $\approx$10,000
objects in a $16'.4\times 16'.4$ field centered on NGC 4472, and
presented the analysis of the metallicities of about 1800 bright ($T_1<23$ mag)
globular clusters. Our most striking result was 
that the metallicity distribution of the globular
clusters  was clearly bimodal, showing distinct peaks at 
[Fe/H] = --1.3 and --0.1 dex. 

In this paper we examine the globular cluster luminosity function (GCLF) and
spatial structure of the NGC 4472 GCS. 
This study is based on data for a large number of globular cluster candidates
in NGC 4472 obtained from deep wide field CCD photometry, while previous
studies on this subject were based either 
on shallow wide field photographic photometry or on deep small field CCD photometry
(\cite{har78}, \cite{har81}, \cite{har86}, \cite{coh88}, \cite{haret91}, \cite{cou91}).

The use of the turnover of the GCLF has become increasingly important as a 
distance indicator. Jacoby \etal (1992) have shown that this technique is more
precise  than previously 
regarded. The peak of the GCLF in several Virgo galaxies
has been used to derive their distances 
(e.g. \cite{haret91}, \cite{whi95}, \cite{els96}).
Our large sample should allow us to obtain a 
more robust value than derived from previous smaller-scale studies. Secondly, 
Ashman \etal (1995) have predicted a metallicity dependence of the GCLF peak,
with metal-poor clusters peaking at brighter magnitude than metal-rich clusters, with the 
exact values depending on the metallicity and bandpass. Whitmore \etal (1995)'s HST
data on the M87 GCS showed a bimodal metallicity distribution and the GCLFs of the two
populations showed peaks which differed by about the amount predicted by 
Ashman \etal. The presence of another bimodal GCS allows another 
critical test of the Ashman \etal prediction, as cluster populations 
with significantly different metallicities can be compared directly.

The utility of a global measure of the globular cluster population in a galaxy,
namely the specific frequency $S_N$, has been shown by many studies (e.g. 
\cite{har91}). 
This provides a luminosity-independent means of comparing the cluster
formation efficiencies of different galaxies, a key ingredient in understanding
how the clusters and the galaxies themselves are made. Likewise, the spatial
structure of the GCS and how it compares to the galaxy light has been used 
since the seminal study of Strom \etal (1981) to set constraints on the 
relative formation and chemical evolution histories of these components. The 
presence of two distinct GC populations in NGC 4472 allow us to investigate their
specific frequencies and spatial structures independently. Such investigations 
are unprecedented.
Note that HST studies are generally confined to only 
a very small inner region of the GCS, while 
wide-field ground-based studies such as ours 
cover virtually the entire extent of the GCS {\bf except}
for the inner regions, which
are saturated and/or subject to crowding and high background light levels.
Thus, HST and ground-based studies are very complementary.

This paper is organized as follows.
Section 2 describes briefly the data for the globular clusters used in this
study. Sec. 3 derives the GCLF and
estimates the distance to NGC 4472.
Sec. 4 estimates the specific frequency of the GCS
in NGC 4472 and investigates its variation with 
galactocentric radius.
Sec. 5 studies the spatial structure of the GCS
and compares this with the structure of the underlying halo starlight of NGC 4472.
Sec. 6 summarizes our results and compares them with the predictions
of various models of the formation
of globular clusters in giant elliptical galaxies.
Preliminary results of this study were presented in several conference
proceedings (\cite{lee96}, \cite{kim96}).

\section{THE DATA}

We have used the photometric data of the objects in 
a $16'.4\times 16'.4$ field centered on NGC 4472 given in Paper I.
The KPNO 4m PF/CCD system was used to obtain deep Washington system (\cite{can76})
$CT_1$ observations, and the data were reduced with the DAOPHOT II 
psf-fitting reduction program.
Figure 1 displays the color-magnitude diagram of the $\approx$10,000
measured objects.
The remarkable bimodal vertical structure graphically portrays
the two populations of
globular clusters in NGC 4472, and the faint blue horizontal structure represents
mostly background galaxies, as described in detail in Paper I.
Here, we have chosen, as the best globular cluster candidates, 
those objects with colors of  $1.0<(C-T_1)<2.3$ 
for investigating the GCLF, going down to our
incompleteness limit of \T1$=24.6$ (see below), and 
have used the bright globular clusters, with $19.63<T_1<23$ mag for
the analysis of the spatial structure of the GCS, 
as marked in Figure 1. These limits (other than the completeness limit determined
below) were derived in Paper I from a careful analysis of the color, magnitude
and spatial distribution of our entire sample of objects. We have relaxed the
photometric error requirement for this analysis to $\sigma(C-T_1)<0.15$.
This increases our final sample of GC candidates to $\sim 2000$.
 
Since the color distribution of the globular clusters in NGC 4472 is clearly
bimodal with a minimum at $(C-T_1)\sim 1.65$,  and an age difference sufficiently
large to explain the observed color difference is very unlikely (Paper I), 
we assume the colors reflect metallicity and have divided the final sample 
into two groups accordingly:
the blue globular clusters ($1.0<(C-T_1)<1.65$, 1257 objects) 
and the red globular clusters ($1.65<(C-T_1)<2.3$, 764 objects) 
(see Paper I). The mean metallicities of each of these two groups are,
respectively, [Fe/H] = --1.3 dex and --0.1 dex.  
These two groups are referred to as BGCs (blue or metal-poor GCs) and 
RGCs (red or metal-rich GCs), respectively, hereafter.
The number ratio of the metal-rich GCs to metal-poor GCs is 0.6.
Inspection of the color-magnitude diagram and luminosity function in 
the next section indicates that the contamination due to background
galaxies and foreground stars is minor (less than 10 \%) in the final sample
of the bright globular clusters with $T_1<23$ mag (see also Paper I).

\section{THE GLOBULAR CLUSTER LUMINOSITY FUNCTION}

\subsection{Completeness Tests}

We have tested the completeness of our photometry using artificial star 
experiments with the aid of the ADDSTAR routine in DAOPHOT II. 
We have added 20,000 artificial stars into 10 copies of the original images 
(2,000 stars per image), 
and have estimated the completeness of our photometry by calculating the
recovery probability of the added stars.
Table 1 lists the completeness as a function of magnitude.
The completeness
in Table 1 is defined as $f(C, T_1)$ = N(recovered stars) / N(added stars).

The incompleteness varies depending on position in the image, and is 
especially large
in the central region  because
of the saturation      due to the bright nucleus of NGC 4472, and in the outer
region due to image quality degradation from the large, poorly corrected
4m PF/CCD field.
Therefore we have limited our final sample to the globular clusters
located within the galactocentric radius range of $50''<r< 420''$
for the following analysis. The lower limit has been increased over that used in
Paper I to ensure uniform
completeness over the range.

Table 1 shows that our photometry is almost complete up to $T_1\sim 24$ mag
and $C\sim 25$ mag,
and the limiting magnitude, defined as the 50\% completeness level,
is $T_{1,lim} \sim 24.6$ mag and $C_{lim} \sim 25.9$ mag.
Table 1 also lists the mean differences of the magnitudes of the added 
and recovered stars, which shows this quantity is generally small until one
approaches the limiting magnitude. We also include the mean photometric errors
(the rms of the differences).
These values agree well with the internal errors returned by DAOPHOT.
 
\subsection{ The Globular Cluster Luminosity Function}

We have derived the luminosity function of this sample and 
the results are shown in Figure 2 and listed in Table 2a.
Before discussing these results, we must first note the limitations of 
our data.
The area covered by this sample is only 152 square arcmin,
although the CCD field covered 
269 square arcmin and reaches $r>8\arcmin$ from the center of the galaxy.
Unfortunately,
we have had to limit our sample to this smaller radial range because of the 
increasingly poor image quality beyond this. Therefore, we are only able to
study the GCS from $\sim 1\arcmin -7\arcmin$. 
Indeed, due to this same problem, we are
unable to derive any measure of the background contamination since we are
unable to perform useful photometry in a region far enough from the galaxy
to lie beyond the extent of the GCS (note that
Harris 1986 estimates the GCs to 
extend out to $r\sim 20\arcmin$) despite our efforts to do so.
We were also unable to observe a separate
comparison field due to time constraints.
In addition, the strongly varying image quality prohibited us from performing 
the usual image classification analysis. We attempted to use the various image
moments defined by Harris \etal (1991)
 to differentiate GCs from resolved background
galaxies, but our attempts proved unsatisfactory. 
Therefore we could not discard resolved objects from our analysis or
subtract the
contribution due to field populations from the derived luminosity function
of the globular clusters. Instead we have investigated roughly the contamination
effect on the resulting luminosity function of the globular cluster candidates
as follows.

In Figure 2(a) and 2(b) we have plotted the luminosity function
of the blue objects ($0.2<(C-T_1)<0.8$), which are 
mostly background  galaxies (the dashed line).
We have chosen this color range to avoid any contamination effect due
to the faint blue globular clusters in deriving the luminosity function 
of the background galaxies.
Figure 2(a) and (b) show that the luminosity function of the blue galaxies is 
negligible up to $T_1\sim 23$ mag, 
but increases rapidly faintward from $T_1\sim23.2$ mag, 
showing a turnover at $T_1\sim 24.9$ mag. 
which  is close to the magnitude limit of our photometry. 
This turnover is undoubtedly      due to the incompleteness 
of our photometry.
It is of course impossible to distinguish individually globular clusters 
from unresolved background
galaxies  within our GCS color and magnitude limits.
The color-magnitude diagram of faint field galaxies ($T_1 \gtsim 23$ mag)
 in other fields shows that their color range 
is typically  $-0.2<(C-T_1)<1.6$. 
Therefore it is concluded 
(a) that the contribution due to the field populations is, at most, minor,
in the sample of the bright ($T_1<23$ mag) globular clusters, 
and in the sample of the RGCs up to the faint magnitude limit;
but (b) that the field contamination starts to become
significant faintward from $T_1\sim 23.2$ mag in the BGCs.

Figure 2(a) and (b) display the luminosity function of the globular cluster candidates
 with $1.0<(C-T_1)<2.3$.
The fainter part ($T_1 > 23.8$ mag) of this luminosity function is clearly 
affected
by the contamination due to the background galaxies, as shown
by the asymmetry around the peak. 
The contamination effect due to the background
galaxies is reduced in the sample from which the bluest objects were removed.
The luminosity functions of the objects with $1.3<(C-T_1)<2.3$ 
and the RGCs in Figure 2(a) and (b) illustrate this effect very well.
Therefore, we have limited our GCLF analysis to objects brighter than 
\T1$=23.8$ with colors of $1.3<(C-T_1)<2.3$.
Using these criteria,    we feel that our data is 
sufficient to investigate the turnover in the GCLF, which lies at a bright 
enough magnitude that the various limitations are tractable.

Even though the luminosity functions of the various GC samples in Figure 2(b)
show somewhat different shapes at the
faint end, all three show clearly a peak at the same magnitude of $T_1=
23.3\pm0.1$ mag.  
The bright part of the derived luminosity function of the globular clusters
is well fit by a Gaussian function with a peak at $T_1=23.3$ mag and
a width $\sigma = 1.3 $, as shown in Figure 3.
For this exercise, we fixed the width at a value typical of those found in 
giant elliptical GCSs
(e.g. Whitmore \etal 1995) since we had the smallest leverage on this
quantity, and derived a formal value of $23.31\pm0.07$ for the peak. Fixing the
width at 1.4 increased the peak by only 0.03 mags.
 
Previous studies of the luminosity functions of the globular clusters in NGC 4472
were based 
on deep small field photometry (\cite{coh88}, \cite{haret91}).
Cohen (1988) presented the $g$-band luminosity function of the globular clusters
located in the radial range of $30''<r<300''$, and
Harris \etal (1991) presented $B$-band luminosity
functions of globular clusters in a $4'.1 \times 3'.4$ field centered on the
nucleus of NGC 4472. 
We have compared these with ours in Figure 2(c).
We have applied  rough color relations appropriate for these globular clusters
to convert $g$ and $B$ magnitudes into $T_1$ magnitudes 
for the purpose of comparison:
$T_1 \sim B-1.2$ and $T_1 \sim g-0.55$ (\cite{coh88}, \cite{djo93}).
Figure 2(c) shows that the three sets of luminosity functions agree
well in general, except for the fact that Cohen's shows some excess
in the range of $23<T_1<24$ mag compared with the other two.
However, ours are based on a much larger number of
globular clusters  and deeper photometry than the others, but do suffer from 
the inability to perform image classification and background analysis.
Also note that the Ajhar \etal (1994) data on NGC 4472 reveal a turnover at
$R(\approx T_1)\sim 23.3$, in excellent agreement with our result.

\subsection{ The Peak Luminosity Difference between the BGCs and RGCs}

We have investigated whether there is any systematic difference in the peak value
of the luminosity functions of the BGCs and RGCs.
The luminosity functions of the BGCs and RGCs are given in Table 2b and
 displayed in Figure 3.
For the BGCs in Figure 3 we have counted only globular clusters with colors of 
$1.3<(C-T_1)<1.65$ instead of $1.0<(C-T_1)<1.65$  
to reduce the field galaxy contamination at the faint end.
The number of globular clusters in this sample is similar to that
of the RGCs so that the two samples can be compared directly on a statistical
basis. We have corrected these samples for the \T1 incompleteness but not for
$C$ incompleteness, and therefore it is likely that we are losing some of the 
reddest, faintest clusters.
Figure 3 shows (a) that the shapes of the bright part of the luminosity function 
of the BGCs and RGCs are similar;
(b) that the faint part of the luminosity function of the BGCs is more 
significantly affected by faint galaxies than that of the RGCs; and 
(c) that both the luminosity functions of the BGCs and RGCs show a peak
at the same magnitude, $T_1=23.3$ mag. We are unable to quantify rigorously
the similarity of these two peaks due to the various problems but estimate that
they are the same to within $\sim 0.15$ mag.

Therefore it appears that there is little, if any, difference in the peak
luminosities of the GCLFs of the BGCs and RGCs.
This result is not consistent with the theoretical prediction by
Ashman \etal (1995). 
According to the relations between peak luminosity vs. metallicity 
given by them,         the peak luminosity
of the RGCs should be 0.3 mag fainter in \T1 than that of the BGCs 
in the case of NGC 4472, which is not seen in Figure 3. 
This difference should be even larger in the more metallicity sensitive $C$ filter.
 We have investigated this effect but are unable to make any useful comparison
due to the data limitations.

In the case of M87, the cD galaxy in Virgo, 
the observational results published so far are somewhat confusing.
Elson \& Santiago (1996) suggested from the study of
254 globular clusters in a field $2'.5$ from the center of M87 
that the peak luminosity of the BGCs is 
0.3 mag brighter than that of the RGCs.
However, this conclusion is based on a small number of globular clusters.
On the other hand, Whitmore \etal (1995) found only half this difference 
between the two populations from their study of 1032
globular clusters in a field $<1'.7$ from the center of M87.
The latter is consistent with our results on NGC 4472.

Finally, we note a curious feature seen clearly in Figure 1: in the brightest
 magnitude of the GCLF there are about twice as many BGCs as RGCs. 
This feature appears to be stronger than expected from the differences in
population sizes but requires more study.

\subsection{ The Distance to NGC 4472}

We have estimated the distance to NGC 4472 using the peak value
of the luminosity function of the globular clusters obtained in the previous
section, $T_1 = 23.3\pm0.1$ mag,
in comparison with the luminosity function of the globular clusters
in our Galaxy. 
Figure 2(d) displays the $R$-band luminosity function 
of the Galactic globular clusters
which is derived from the compiled data of 
Djorgovski \& Meylan (1993) and Harris (1996). 
We have plotted this figure in such a way that the position of the peaks in the
GCLFs for NGC 4472 and our Galaxy are
at the same location in the horizontal scale.
We have adopted the metallicity-luminosity relation of the RR Lyraes of
$M_V= 0.17[{\rm Fe/H}] + 0.82$, as discussed in Lee \etal (1993), 
in deriving the absolute magnitudes of the Galactic globular clusters.
In Figure 2(d) we have plotted the luminosity functions for two samples:
one for the halo globular clusters, the other for both halo and disk globular
clusters. 
The luminosity functions of both samples of the Galactic globular clusters show clearly a peak
at $M_R=-7.9\pm 0.1$ mag (the difference between $R$ and $T_1$ magnitudes
is negligible (\cite{gei96a})).

Comparing these peak values of the GCLFs 
we derive a distance estimate of 
$(m-M)_0=31.2\pm 0.2$ for zero reddening (\cite{bur82}), 
which corresponds to a distance of $17.4\pm 1.6$ Mpc.
This distance estimate is consistent with the value based 
on the $B$-band GCLF (\cite{haret91}),
 $(m-M)_0=31.3$, and
the estimate based on the surface brightness fluctuation method,
$(m-M)_0=31.18$ (\cite{ton90}). However, our estimate
is much larger than the value derived using the planetary nebula luminosity
function, $(m-M)_0=30.7 \pm 0.19$ (\cite{jac90}).
Note that Ciardullo \etal (1993) presented later, in a comparative study of the
planetary nebulae luminosity function 
and surface brightness fluctuation distance scales,
updated values based on these two methods for NGC 4472: 
$(m-M)_0=30.84\pm0.11$ and $30.78\pm0.07$,
respectively, which are significantly smaller than ours.

Our distance estimate for NGC 4472 is very similar to that for M87,
the cD galaxy in the core of Virgo, $\approx 4^\circ$ away to the north
of NGC 4472 (\cite{lee93a}, \cite{lee93b}, \cite{whi95}). 
Lee \& Geisler (1993b) estimated, from deep Washington CCD photometry, 
a peak GCLF value of $T_1=23.4 \pm0.2$ mag 
(corresponding approximately to $V=23.85$ mag).
This estimate was confirmed later by more comprehensive data from Whitmore 
tal (1995)
based on Hubble Space Telescope observations of over 1000 globular clusters
in the central region of M87, who derived 
$V_{TO}=23.72\pm 0.06 $ mag (\cite{whi95}).
Considering the foreground reddening of $E(B-V)=0.02$ for M87 (\cite{bur84})
and adopting the same peak luminosity of the Galactic globular clusters as used
for NGC 4472,
we obtain an estimate for the distance modulus of M87 from the HST data,
$(m-M)_0=31.2\pm 0.1$. This value is slightly larger than the data from
Whitmore \etal (1995), because they adopted an 0.1 mag fainter
value for the peak luminosity of the Galactic globular clusters than that
used in this study.       

These results show that NGC 4472 and M87 are basically at the same distance
from us, $d=17.4\pm 1.6$ Mpc, even though they are $4^\circ$ apart in the sky.
Adopting this value for the distance to Virgo yields estimates
for the Hubble constant, 
$H_0=68\pm 6$ \kms \, for the Virgo velocity of $1179\pm 17 $ \kms (\cite{jer93}), 
or  
$H_0=81\pm 9$ \kms \, for the Virgo velocity of $1404\pm 80$ \kms (\cite{huc88}).

\section{SPECIFIC FREQUENCY}

The specific frequency, $S_N$, of the GCS of a galaxy
is the total number of globular clusters ($N_{\rm tot}$)
per unit galaxy luminosity  
($M_V=-15$ mag), defined as $S_N=N_{\rm tot} 10^{0.4(M_V^T+15)}$
(\cite{har81}). It represents the global efficiency of globular cluster
formation over the history of the parent galaxy.
We have estimated the specific frequency for NGC 4472 
using the luminosity functions
of the globular clusters given in the previous section as follows.

First, we calculate the total number of clusters which would have been observed
in the CCD images, if the observations were deep enough to cover the faintest
clusters.
Assuming that the luminosity function is symmetric around the peak, 
we estimate the total number of globular clusters by doubling
the number of observed clusters brighter than the peak luminosity,
obtaining a value of 4116 clusters.
Secondly, we estimate the correction due to lack of areal coverage for
  the central ($<50''$) area and outer ($r>420''$) area of NGC 4472,
using the power-law and deVaucouleurs law for the radial profiles of the
surface number density of the globular clusters
obtained in the following section. We extrapolate the observed profiles
out to $10'$.

We next estimate roughly the background level in our data considering 
the following: 
(a) The background level for NGC 4472 given by Harris (1986) is 4.3 objects per 
square arcmin for the magnitude limit of $V=23.3$ mag (corresponding to 
$T_1\sim 22.8$ mag);
(b) In creating the globular cluster sample with colors $1.0<(C-T_1)<2.3$ 
we have already subtracted the non globular cluster candidates 
with colors $(C-T_1)<1.0$ or $>2.3$. 
Therefore we need to subtract only the field objects within the color range of
the selected globular clusters from the globular cluster sample.
Considering these two points, we derive roughly 
2.4 objects per squarearcmin for the
background level in our data.
However, this value is much higher than the number density for the very outer
area of the field ($500''<r<550''$) where some globular clusters are included.
Therefore this value is considered as an upper limit to the background level.

Finally we derive the fully corrected total number of globular clusters in NGC 4472,
obtaining $6700\pm 600$ and $5500\pm 500$ for the power-law and de Vaucouleurs law, respectively.
The mean of these two values is $6100 \pm 800$, which is
adopted as $N_{tot}$ for the NGC 4472 GCS.

The total $V$ magnitude of NGC 4472 is $V^T=8.41$ mag (RC3), 
which corresponds to an absolute magnitude of  $M_V^T= -22.8$ mag 
for the distance determined in the previous section, $d=17.4\pm 1.6$ Mpc.
From the total number of globular clusters and the absolute magnitude of NGC 4472
we derive a value for the specific frequency, $S_N=4.7 \pm 0.6$.
A change in the distance modulus of +0.1 corresponds to a decrease in
the specific frequency of -0.4.

Harris (1986, 1991) estimated the total number of globular clusters in NGC 4472
to be $7400\pm2100$, from the observed number of clusters with $V<23.3$ mag,
$2300\pm 400$. He derived a specific frequency, $S_N=5.0\pm 1.4$ from these
data for $(m-M)_0=31.3$. 
Our results are consistent with Harris' within the errors of the estimates,
but ours have smaller errors than Harris' and are derived from a larger 
sample of clusters.

We have  investigated the radial variation of the local specific frequency,
as shown in Figure 4, which is also listed in Table 3. 
The surface photometry of the halo used in this diagram is described in Kim \etal (1997).
Figure 4 also displays the results given by McLaughlin \etal (1994)
which were based on integrating the power law obtained from the luminosity 
function data of Harris (1986). McLaughlin \etal adopted a value for the
distance modulus of $(m-M)_0=31.0$, which is 0.2 smaller than ours.
Our results are consistent with McLaughlin {\it et al.}'s
for an intermediate distance modulus, except in the 
outer regions, where our values fall substantially below his. 
McLaughlin {\it et al.}'s $S_N$ values continue to increase even more than linearly with 
distance out to at least $20\arcmin$, reaching a value of $\sim 32$ at this
radius.  But note that this result is based on photographic GC and surface 
photometry data, and the latter  only extend to
$10\arcmin$. Given these limitations, we feel that the McLaughlin {\it et al.}'s
specific frequency values beyond $\sim 8\arcmin$ are very uncertain, and in the outer parts
represent a very large extrapolation.
We find, instead, that $S_N$ increases linearly with distance only
out to $\sim 5.5\arcmin$, beyond which $S_N$ levels off with a value of only
about 8.5, some 4 times lower than the value derived by McLaughlin \etal for
the outermost regions. Although our radial coverage does not extend beyond $7
\arcmin$, it is difficult to believe that the local $S_N$ would begin to increase
again at some point. Note that McLaughlin {\it et al.}'s actual data for M87 shows 
a very similar behavior to that we find for NGC 4472 -- a linear rise with radius
out to about $7\arcmin$ with no further significant increase out to the limit 
of their data at about $9\arcmin$. The implications of these differences are
quite important for some models of GC formation, as will be discussed in Section
6.

Our data allow us to investigate for the first time
how the local specific frequency of the two  GC populations vary with
radius, as also 
shown in Figure 4. We see that both the RGCs and the BGCs increase 
their specific frequency linearly with radius from the central regions out to $\sim 5.5\arcmin$
and then level off with no further significant increase. Both of the populations
have a very similar specific frequency in the inner regions but the BGC specific frequency increases more 
rapidly with radius than that of the RGCs, such that in the outer regions the 
specific frequency of the BGCs is $\sim$ twice that of the RGCs. Thus,
if a galaxy with a GCS similar 
to that of NGC 4472 were subject to stripping, the stripped 
material would have a 
specific frequency $\sim 8.5$ and BGCs would outnumber RGCs by about a factor of 2.

\section{SPATIAL STRUCTURE OF THE GLOBULAR CLUSTER SYSTEM}

\subsection{Overall Structure}

We have investigated the spatial structure of the GCS in NGC 4472 
using the final sample of the bright globular clusters with
$T_1<23$.
Figure 5 displays the spatial positions of these objects.
Two important features are immediately seen in Figure 5:
(1) The globular cluster candidates (5a) are centrally
concentrated, while the non-globular cluster objects 
(5b - 
with the colors of $(C-T_1)<1.0$ or $(C-T_1)>2.3$) are uniformly distributed
over the field;
(2) The spatial distribution of the BGCs (5c) is clearly different from 
that of the RGCs (5d) in the sense that the latter is more spatially concentrated
than the former. We also note a strong clumping of both the BGCs and RGCs in the
inner NW and W directions, respectively. Similar clumpiness
has been seen before in both the M87 and NGC 1399 GCSs (Pritchet 1996) but is
presently not explained.

To investigate in detail the spatial structure of the globular cluster systems
as seen in Figure 5,
we have created globular cluster number density maps by counting globular 
clusters in a  $24''  \times  24''$ reseau.
Figure 6 illustrates the resulting cluster density
contour maps.
Note that the central region of $3\times3$ reseaus in Figure 6 is masked 
because of the saturated nucleus of NGC 4472. 
Figure 6(a) and (b) show that the spatial structure of the entire GCS is
almost spherical in the outer area and somewhat elongated in the inner region, 
while the structure of the halo starlight of NGC 4472 is significantly 
elongated overall (with ellipticity $\sim 0.2$ and position angle 
$150^\circ-160^\circ$).
Thus the overall structure of the entire GCS is
different from that of the halo starlight.

Previously Harris \& Petri (1978), 
from a photographic study of a region with $2\arcmin<r<20\arcmin$,
suggested that the GCS in NGC 4472
is almost spherically symmetric, while Cohen (1988) argued (from CCD data in a
region with $0.5\arcmin<r<5\arcmin$) that
the GCS has the same ellipticity and 
position angle as the halo light.
Our result indicates that the opposite conclusions given by these two studies
are probably due to the mostly different radial ranges of NGC 4472
studied by them. 

In addition, Figure 6(c) and 6(d) show striking evidence
that the spatial structures of the BGCs and RGCs are
 systematically different: 
(1) The RGCs are more centrally concentrated than the BGCs; 
(note that the contour levels are the same in Figure 6(c) and (d)). 
(2) The spatial structure of the RGCs is significantly elongated 
roughly along the major axis of the halo of NGC 4472, 
while that of the  BGCs is almost spherical.
With a large sample of globular clusters in a wide field, we find 
for the first time 
that there are systematic differences in the spatial structure 
of the BGCs and RGCs in a giant elliptical galaxy. This is further strong 
evidence for the existence of two distinct populations of GCs in such a galaxy.
Quantitative analysis of the spatial structure of the GCS 
follows.

\subsection{Radial Structure}

We have determined the structural parameters 
of the RGCs and BGCs, 
fitting ellipses to the number density maps of the globular clusters 
created as above. 
We have used the ellipse fitting routine ELLIPSE in IRAF/STSDAS for this purpose.
Figure 7 displays the variation of the surface number density, color, ellipticity,
and position angle with respect to the projected galactocentric radius 
for the two populations, and these results are listed in Table 4.

\underline{Surface number density:}
Figure 7(a) shows the radial profiles of the surface number density of the globular
cluster systems we have derived.
In Figure 8(a) we have compared our results for the entire GCS
with Harris(1986)'s based
on photographic counts of a wide field and Harris \etal (1991)'s
based on CCD photometry of a small field close to the center of NGC 4472.
In Figure 8(a) we have plotted the globular clusters brighter than
$T_1<22.85$ mag, which is the limiting magnitude of the data in Harris (1986),
and corrected our data for the background level derived above.
These figures show that our results are roughly consistent with the other two
in the overlapping radial range.
We have fit the radial profiles of the entire GCS with $T_1<22.85$ mag 
using a power law and a deVaucouleurs law: 
$\log \sigma_{GC} = (1.60\pm0.04) - (1.23\pm0.06) \log r$(arcmin) and  
$\log \sigma_{GC} = (3.16\pm0.12) - (1.62\pm0.08) r^{1 \over 4}$ 
for the range of $70''<r<400''$. 
These  coefficients    are not significantly affected if we 
change from  the $T<22.85$ mag sample to the $T_1<23.0$ mag sample.
In addition, Figure 7(a) shows that
the surface number density of the RGCs decreases faster than that
of the BGCs.
We have fit the radial profiles of the BGCs and RGCs based on the original sample
with $T_1<23.0$ mag. (i.e. uncorrected for background contamination),
using the deVaucouleurs law and a power law, as shown in Figure 8(b):
$\log \sigma_{GC} = (2.43\pm0.15) - (1.13\pm0.10) r^{1 \over 4}$
(and  $= (1.36\pm0.05) - (0.88\pm0.08) \log r$) for the
BGCs and
$\log \sigma_{GC} = (3.05\pm0.12) - (1.72\pm0.09) r^{1 \over 4}$
(and $= (1.39\pm0.04) - (1.28\pm0.07) \log r$) for the RGCs.
We have no information on the background levels for the BGCs and RGCs.
In order to make a crude first-order correction,
we assume the ratio of the background level for the BGCs and RGCs is one, and
use the rough value for the background levels for $T_1<22.85$ mag sample.
Figures 7(a) and 8(c) show the radial profiles
after subtracting the  background level from the number density for the BGCs and RGCs.
Fitting similarly to these data,
we obtain
$\log \sigma_{GC} = (2.75\pm0.17) - (1.42\pm0.12) r^{1 \over 4}$
(and $= (1.40\pm0.06) - (1.12\pm0.09) \log r$) for the
BGCs and
$\log \sigma_{GC} = (3.66\pm0.19) - (2.28\pm0.14) r^{1 \over 4}$
(and $= (1.47\pm0.06) - (1.73\pm0.11) \log r$) for the RGCs.
Thus, the two populations have a spatial
structure that differs at the $\sim 6\sigma$ level.

\underline{Color:}
Figure 7(b) shows that the median colors of the entire GCS
show a strong radial gradient, while the RGCs and BGCs individually each
show little, if any, radial gradient. 
The slopes ( = d[Fe/H]/d$\log r$(arcmin)) determined in Paper I are
$-0.41\pm0.03$  for the entire GCS, $-0.15\pm0.03$ for the
BGCs, and $-0.12\pm 0.06$ for the RGCs.
This indicates that the large radial gradient of the median colors
of the entire GCS is simply due to the varying radial
mixture of the two   populations,
rather than to dissipative processes, as discussed in detail in Paper I.

\underline{Ellipticity:}
Figure 7(c) shows that the mean ellipticity of the entire GCS is 
almost constant around the value of 0.16. 
The mean ellipticity of the RGCs increases from $\approx 0.06$ 
continuously outward up  to $\approx 0.3$ at $r=410''$.
On the other hand, the ellipticity of the BGCs is
almost constant around the value of 0.07, except for the range of
$150''<r<240''$ where the ellipticities are as large as 0.27. The
ellipticity for this range varies abruptly from nearby values. 
This variation is considered to be due to some local enhancement of the
number density in the south east direction, as seen in Figure 6(c).   

\subsection{Azimuthal Structure}

We have investigated the azimuthal distribution of the globular clusters.
Figure 7(d) shows that the mean position angles of the entire GCS and
the RGCs are similar with the value of about $150^\circ$.
The position angles of the BGCs vary from $\sim100^\circ$ to
$\sim180^\circ$. The significance of this large variation is small, considering
the difficulty of accurate determination of the position angles for small
ellipticities, as in this case.
Figure 9 displays azimuthal variations of the globular cluster number density
for the galactocentric radius of $70''<r<420''$.
It shows that the entire GCS has an obvious peak 
at the position angle of $\approx 155^\circ$, 
and that the RGCs have a peak at the same value, 
while the BGCs show an almost uniform distribution.
The presence of only this one peak instead of two in Figure 9
is due to the asymmetric distribution of the globular clusters extended
along the southeast direction. 

\subsection{Comparison of the Globular Cluster Populations with the Halo Light}

We have compared the structural parameters of the globular cluster populations with
those of the halo starlight of NGC 4472. The galaxy surface photometry
is given in  Kim \etal (1997). Note that our new results for the outer halo have
changed significantly from those given in Paper I,
due to a small change in the adopted $C$ and \T1 background levels.
Figure 7 reveals significant features as follows:
(a) The radial profiles of the number density of the GCS 
are flatter than that of the halo, being spatially more extended than the halo.
Note that the BGCs are significantly more extended than the halo, while
the RGCs are marginally more extended than the halo. 
The $T_1$ surface brightness profiles are fit using the deVaucouleurs law 
and a power law,
$\log \mu_{T_1} = (13.89\pm0.01) + (6.33\pm0.01) r^{1 \over 4}$
and $\log \mu_{T_1} = (20.08\pm0.00) + (4.59\pm0.01) \log r$ (\cite{kim97}).
This slope of the halo is slightly steeper than the slope of the RGCs
based on the background subtracted sample. However, the background level
used for the RGCs is considered to be an overestimate, discussed in Section 4.
Better estimation of the background level based on wider field imaging data
is needed to obtain a more reliable estimate of the slope of the RGCs;
(b) The median color of the RGCs is remarkably similar to   
that of the halo, while the median color of the BGCs is
much bluer (by $\approx$0.5) than that of the halo;
(c) The mean ellipticity of the entire GCS is slightly smaller than
that of the halo. The ellipticity of the RGCs becomes larger than
that of the halo in the outer region,   $r>300''$, while the ellipticity
of the BGCs is generally much smaller than that of the halo;
(d) The mean position angle of the entire GCS is similar to that of the
halo for $r>200''$.
In general, we see that the RGCs are more similar 
to the halo stars of NGC 4472 in terms of spatial structure and colors
than are the BGCs.

\section{SUMMARY AND DISCUSSION}

Although it is risky to test galaxy formation models based on a single
case, the fact is that our study of the NGC 4472 GCS represents the best data
yet available which combines both deep widefield photometry, to ensure large
samples and large radial coverage, as well as a wide color baseline that allows
high metallicity precision.  In particular, we have large numbers of both RGCs and BGCs
with which to study their properties individually for the first time.
Thus, although we strongly urge similar studies (and are
carrying out such studies), we will examine the NGC 4472 GCS in some detail and 
compare our results to several GCS and galaxy formation models.

First, we summarize the salient observed characteristics of the
NGC 4472 GCS as found in Paper I and this study: 

\begin{enumerate}

\item The median metallicity of the entire GCS is [Fe/H]$=-0.9\pm0.2$ dex. 

\item The GCS consists of two distinct
populations: one metal-poor ([Fe/H] $\sim -1.3$ dex) and one metal-rich 
([Fe/H] $\sim -0.1$ dex), with metallicity spreads in each population consistent
with that of the halo GCs in our Galaxy ($\sigma$ ([Fe/H])$\sim 0.35$ dex).  

\item The number ratio RGC/BGC = 0.6.

\item The luminosity function of the globular clusters shows a peak
at $T_1 = 23.3 \pm 0.1$ mag.

\item The total number of globular clusters in NGC 4472 within $r\sim10'$
is estimated to be $6100\pm800$, 
and the global specific frequency is $S_N= 4.7\pm0.6$. The local
specific frequency increases linearly
with radius from $\sim 1.5$ at $r=1\arcmin$ to $\sim 8.5$ at $r\sim
5.5\arcmin$ but does not increase further beyond this distance, to at least 
$7\arcmin$. Both the RGCs and 
the BGCs show similar radial trends, with the local specific frequency of the 
BGCs increasing more rapidly with radius
such that in the outer regions there are about twice as many BGCs as RGCs.

\item The RGCs are 
more centrally concentrated than the BGCs.
The BGCs are significantly more extended than the halo, while
the RGCs are marginally more extended than the halo 

\item The metallicity gradient of the entire GCS
is $\Delta{\rm [Fe/H]} / \Delta \log r = -0.41\pm0.03$, but most
of this gradient appears to be due to the varying radial concentration
of the two populations, and not to dissipative processes, since neither
the RGCs or the BGCs show significant gradients.

\item The color of the RGCs is remarkably similar to that of the halo
starlight, while the color of the BGCs is much bluer than
that of the halo star light, indicating a metallicity difference of
[Fe/H]$\sim 1.2$ dex.

\item The spatial structure of the entire GCS
is almost spherical in the outer region and somewhat elongated in the
inner region.
The BGCs are distributed almost spherically,
while the RGCs are elongated
along the major axis of the halo star light.  

\end{enumerate}

These results can be used as constraints on models for the origin of globular
clusters. There have been several models  suggested,
a brief summary of which is given in Ashman \& Zepf (1997) and
Forbes \etal (1997 - hereafter FBG97).
Here we consider only two models:
the gaseous merger model (\cite{ash92} - hereafter AZ92, \cite{zep93}, see also
\cite{her93})     
and the episodic {\it in situ} formation plus tidal stripping model (FBG97).

In the gaseous merger model, an elliptical galaxy is formed by the merging of 
two or more gas-rich spiral galaxies. 
The spiral galaxies have enriched gas in the disk and metal-poor globular
clusters in the halo. 
New globular clusters are formed from the enriched gas 
during the merging/interacting process.
The resulting elliptical galaxy then has two GC populations:
the younger, more metal-rich, and spatially more concentrated clusters formed
as a result of the merger, and the original, metal-poor, more spatially extended
GCs formed in the progenitor spirals. 
 It is generally true that typical cluster
ellipticals have a \gtsim factor of 2 higher specific frequency than spirals. (However, we note the need for more complete studies of even nearby galaxies, 
using large area detectors in order to study the entire GCS to fainter magnitude 
levels than in previous studies, and thus improve existing $S_N$ values.)
In order to account for the \gtsim factor of 2 higher specific frequency
of ellipticals compared to spirals, which is a key idea of the merger model
as proposed by AZ92, at least
as many new RGCs are needed to be made in the merger as 
were originally present in the combined total of BGCs in the progenitor spirals.

This merger model has been very successful in explaining several observed
qualitative
characteristics of the globular clusters in giant elliptical galaxies and
interacting galaxies: for example, the presence of young massive clusters
in interacting galaxies (e.g. Whitmore \& Schweizer 1995),
the bi(or multi)-modal metallicity distributions
in a number of GCSs (Paper I, FBG97), 
and the lack of metallicity gradients in either of the two populations but the
presence of a gradient in the overall GCS (e.g. Paper I).
However, some of the quantitative
predictions of the AZ92 model have not been born out by recent observations. 
This was recently pointed out by FBG97. In
particular, the number of RGCs found in most multimodal ellipticals known is
insufficient to explain the specific frequency differences between typical spirals and ellipticals.
In addition, the global specific frequency is roughly
anticorrelated with the number ratio $N_{RGC}/
N_{BGC}$, opposite to the prediction of AZ92.

To explain such findings, FBG97 proposed another model, 
the episodic {\it in situ} formation plus tidal stripping model. 
In this model metal-poor globular clusters are formed first
at an early stage in the initial collapse of the protogalactic 
cloud with only minor star formation.
After some time ($\approx 4$ Gyrs) of quiescence, 
metal-rich globular clusters are formed
out of more enriched gas, roughly contemporaneously with most of
the galaxy stars,
during the major collapse of the protogalactic cloud, which then forms an
elliptical galaxy with two GC populations.
Most of the globular clusters are formed in the first formation episode and 
should not be structurally related to the halo light, while
the newly formed globular clusters are closely coupled to the
galaxy and share a common chemical enrichment history 
and structural characteristics.

We will discuss each of our above salient points of relevance to these two 
theories in turn:  

Point 2. The existence of multimodal GC populations in giant ellipticals is now 
known to be widespread (Paper I, FBG97)
and perhaps universal. The merger model {\bf predicted} this and is one of its
major successes, while FBG97         use this as an initial theorem from 
which they derive their model. As pointed out by FBG97, the existence of
multimodality rules out simple monolithic collapse models since these would form
only a single population of GCs. However,
the similar metallicity spreads in each population and the constancy of the metallicity difference
between them in different galaxies is 
something that requires both additional observational confirmation (this indeed
applies to all of our points) and 
theoretical explanation. For example, a giant elliptical having suffered
many mergers  would be expected to possess very broad metallicity ranges as opposed
to the distinct, rather narrow peaks that are seen.

Point 3. As in our data on NGC 4472, of the $\sim 13$ well studied giant ellipticals to date
(FBG97, Geisler \& Lee 1998, Lee \& Geisler 1998), only two have $N_{RGC}/
N_{BGC}\gtsim 1$, as required by the merger model. Most cases show a ratio
of only $\sim 0.5$,  a factor of 2 or more lower than needed to explain the specific frequency
differences. This is an obvious blow to the original AZ92 model.

Point 5. The {\it in situ} scenario predicts that higher specific frequency values should be associated
with the BGCs, since they formed at a time when cluster formation dominated 
star formation, and this agrees with our finding of higher specific frequency for the BGCs 
in the outer regions of the NGC 4472 GCS. However, FBG97 also invoke stripping 
of the outer regions of nearby galaxies to account  for the unusually high specific frequency
values associated with some cD galaxies, such as M87, and refer to McLaughlin
{\it et al.}'s (1994) work to state that the outer regions of some ellipticals, and NGC 4472
in particular, have a very high specific frequency, as much as in the outer regions of
M87. However, we have seen that the outer regions of NGC 4472 
probably have a specific frequency about
4 times lower than the value derived by McLaughlin {\it et al.}, which makes it much 
more difficult to build up large specific frequency galaxies from stripping of the outer 
regions of a GCS like that of NGC 4472.

Point 6. Both models predict that the RGCs will be more centrally concentrated 
than the BGCs. 
The {\it in situ} model further predicts that the RGCs should have the same
spatial distribution as the galaxy halo. 
The RGCs are marginally more extended than the halo light.

Point 7. Both models predict a metallicity gradient in the overall GCS due to the varying
concentration of the two populations. The {\it in situ} model would further predict that
the RGCs should also show a gradient due to dissipation, which is not seen in
NGC 4472.

Point 8. Both models successfully predict the close agreement of the metallicity of the RGCs
and the galaxy light, and the offset of these from the BGCs.

Point 9. Both models would predict a roughly spherical distribution of the BGCs.
FBG97 also predict that the RGCs will closely match the ellipticity of the 
underlying galaxy, as seen in NGC 4472. The merger model would also suggest
that these distributions should be similar, but results may depend on details 
of the merger.

FBG97 also predicted 
that the RGCs should show some  rotation, depending on the amount of dissipation,
while the BGCs should have no rotation and
a large velocity dispersion.
Kinematic studies of a large number of the BGCs and RGCs in NGC 4472 are
 required to test these predictions
(Mould {\it et al.} (1990) measured the velocities of 26 globular clusters
 in NGC 4472, but the number of clusters is too small to study
the differences between the metal-poor and metal-rich systems.)
Such studies are indeed now becoming available: e.g. Sharples \etal (1997) have
obtained spectra of 57 GCs in NGC 4472, with roughly equal numbers of BGCs and
RGCs. They find no sign of rotation in either population, and a strong hint that the
BGCs have a higher velocity dispersion than the RGCs.

Thus, we see that both models have some successes as well as failures when 
compared to observations. FBG97 point out several other problems with the 
merger model. However, as they admit, their scenario lacks an explanation for why
star and cluster formation was shut down for several Gyrs and why it was then
reinitiated. In our view, it is now becoming increasingly clear that mergers
are involved in the formation of many giant ellipticals 
and that a natural explanation of the onset of the second generation
of star and cluster formation is provided by mergers. However, it is also 
now quite clear that this second generation of clusters was in general
insufficient to
make up the difference in specific frequency between typical spirals and ellipticals.
Thus, the AZ92 merger model may well explain the second epoch trigger mechanism,
but, ironically, we may still be left with the original problem they tried 
to solve -- how to account for the specific frequency difference.
The question of the origin of the high specific frequency GCSs remains outstanding.

\acknowledgments

D.G. would like to acknowledge
the Department of Astronomy,
Seoul National University, and E. Geisler 
for their kind hospitality, and understanding and patience, respectively,
during his stay there.
This research is supported by
NON DIRECTED RESEARCH FUND, Korea Research Foundation, 1996 (to M.G.L.). 
This research is
supported in part by NASA through grant No. GO-06699.01-95A (to D.G.) from the
Space Telescope Science
Institute, which is operated by the Association of Universities for
Research in Astronomy, Inc., under NASA contract NAS5-26555.

\clearpage


\begin{deluxetable}{ccrr|ccrr}
\tablecolumns{8}
\tablewidth{0pc}
\tablenum{1}
\tablecaption{Completeness of photometry.}
\tablehead{
\colhead{$C$}                & \colhead{$f(C)$}          & 
\colhead{$\Delta C$}         & \colhead{$\sigma_C$}      & 
\colhead{$T_1$}              & \colhead{$f(T_1)$}        &
\colhead{$\Delta {T_1}$}     & \colhead{$\sigma_{T_1}$}}

\startdata

 20.28 &  1.00 & $ 0.002$ & 0.005 & 18.17 &  1.00 & $-0.002$ & 0.000 \nl
 20.78 &  1.00 & $ 0.000$ & 0.004 & 18.67 &  1.00 & $-0.001$ & 0.003 \nl
 21.28 &  1.00 & $ 0.001$ & 0.005 & 19.17 &  1.00 & $ 0.000$ & 0.004 \nl
 21.78 &  1.00 & $ 0.002$ & 0.009 & 19.67 &  1.00 & $ 0.001$ & 0.006 \nl
 22.28 &  1.00 & $ 0.003$ & 0.010 & 20.17 &  1.00 & $ 0.001$ & 0.007 \nl
 22.78 &  1.00 & $ 0.002$ & 0.017 & 20.67 &  1.00 & $ 0.000$ & 0.008 \nl
 23.28 &  1.00 & $ 0.004$ & 0.026 & 21.17 &  1.00 & $ 0.002$ & 0.014 \nl
 23.78 &  0.99 & $ 0.000$ & 0.040 & 21.67 &  1.00 & $ 0.005$ & 0.020 \nl
 24.28 &  0.98 & $ 0.003$ & 0.061 & 22.17 &  1.00 & $ 0.005$ & 0.035 \nl
 24.78 &  0.96 & $ 0.002$ & 0.093 & 22.67 &  0.98 & $ 0.003$ & 0.057 \nl
 25.28 &  0.84 & $-0.002$ & 0.142 & 23.17 &  0.96 & $ 0.006$ & 0.082 \nl
 25.78 &  0.61 & $-0.047$ & 0.207 & 23.67 &  0.95 & $ 0.007$ & 0.128 \nl
 26.28 &  0.33 & $-0.178$ & 0.297 & 24.17 &  0.78 & $-0.002$ & 0.180 \nl
 26.78 &  0.13 & $-0.390$ & 0.316 & 24.67 &  0.47 & $-0.100$ & 0.277 \nl
       &       &          &       & 25.17 &  0.21 & $-0.334$ & 0.311 \nl
       &       &          &       & 25.67 &  0.09 & $-0.638$ & 0.292 \nl

\enddata
\end{deluxetable}

\begin{deluxetable}{cccccccc}
\tablecolumns{8}
\tablewidth{35pc}
\tablenum{2a}
\tablecaption{Luminosity function of all measured globular clusters in NGC 4472.}
\tablehead{
\colhead{$T_1$}     & \colhead{$N(T_1)$}  &
\colhead{$T_1$}     & \colhead{$N(T_1)$}  &
\colhead{$T_1$}     & \colhead{$N(T_1)$}  &
\colhead{$T_1$}     & \colhead{$N(T_1)$} }

\startdata
 19.5 &    6 & 20.7 &   34 & 21.9 &  102 & 23.1 &  195 \nl
 19.7 &   10 & 20.9 &   48 & 22.1 &   99 & 23.3 &  214 \nl
 19.9 &   11 & 21.1 &   62 & 22.3 &  134 & 23.5 &  193 \nl
 20.1 &   11 & 21.3 &   64 & 22.5 &  146 & 23.7 &  165 \nl
 20.3 &   21 & 21.5 &   76 & 22.7 &  168 &      &      \nl
 20.5 &   20 & 21.7 &   94 & 22.9 &  163 &      &      \nl
\enddata
\end{deluxetable}

\begin{deluxetable}{ccccccc}
\tablecolumns{7}
\tablewidth{35pc}
\tablenum{2b}
\tablecaption{Luminosity function of blue and red globular clusters in NGC 4472.}
\tablehead{
\colhead{} & \multicolumn{2}{c}{$N(T_1)$} & \colhead{} &
\colhead{} & \multicolumn{2}{c}{$N(T_1)$} \\
\cline{2-3} \cline{6-7} \\
\colhead{$T_1$}   & \colhead{\sc BGC} & 
\colhead{\sc RGC} & \colhead{}        & 
\colhead{$T_1$}   & \colhead{\sc BGC} &
\colhead{\sc RGC} }

\startdata
 19.65 &    6.0 &    3.3 & & 21.75 &   42.7 &   54.7 \nl
 19.95 &    8.0 &    2.7 & & 22.05 &   50.0 &   50.0 \nl
 20.25 &   10.7 &    7.3 & & 22.35 &   57.1 &   77.9 \nl
 20.55 &   11.3 &   12.7 & & 22.65 &   73.4 &   91.8 \nl
 20.85 &   25.3 &   20.0 & & 22.95 &   81.2 &   89.5 \nl
 21.15 &   39.3 &   25.3 & & 23.25 &  105.0 &  106.4 \nl
 21.45 &   38.0 &   32.7 & & 23.55 &   96.6 &   79.8 \nl
\enddata
\end{deluxetable}

\begin{deluxetable}{cccc}
\tablecolumns{4}
\tablewidth{35pc}
\tablenum{3}
\tablecaption{Radial variaton of the local specific frequency of the globular clusters in NGC 4472.}
\tablehead{
\colhead{R [arcsec]} & \colhead{$S_N$} & 
\colhead{R [arcsec]} & \colhead{$S_N$} }

\startdata
 72 & $1.59\pm 0.09$ & 283 & $5.78\pm 0.34$ \nl
107 & $2.56\pm 0.14$ & 313 & $7.40\pm 0.44$ \nl
137 & $3.77\pm 0.21$ & 339 & $8.60\pm 0.52$ \nl
167 & $4.01\pm 0.23$ & 368 & $8.87\pm 0.55$ \nl
196 & $5.15\pm 0.29$ & 398 & $8.71\pm 0.57$ \nl
224 & $5.20\pm 0.30$ & 433 & $8.66\pm 0.57$ \nl
252 & $5.88\pm 0.34$ &     &                \nl
\enddata
\end{deluxetable}

\begin{deluxetable}{cccccccc}
\tablecolumns{8}
\tablenum{4}
\tablewidth{35pc}
\tablecaption{Radial surface density of the globular cluster system in NGC 4472.}
\tablehead{
\multicolumn{2}{c}{\sc ALL} & \colhead{} & 
\multicolumn{2}{c}{\sc BGC} & \colhead{} &
\multicolumn{2}{c}{\sc RGC} \\
\cline{1-2} \cline{4-5} \cline{7-8} \\
\colhead{R[arcsec]}   & \colhead{$\sigma_{GC}$}   & \colhead{} &
\colhead{R[arcsec]}   & \colhead{$\sigma_{GC}$}   & \colhead{} &
\colhead{R[arcsec]}   & \colhead{$\sigma_{GC}$} } 

\startdata
54 & $31.4\pm 3.1$ & & 61 & $14.6\pm 1.8$ & & 50 & $15.1\pm 2.5$ \nl
97 & $24.3\pm 2.4$ & & 111 & $12.0\pm 1.5$ & & 84 & $15.3\pm 2.6$ \nl
131 & $21.2\pm 2.1$ & & 146 & $11.8\pm 1.5$ & & 111 & $10.2\pm 1.7$ \nl
164 & $15.3\pm 1.5$ & & 181 & $7.6\pm 1.0$ & & 138 & $9.7\pm 1.6$ \nl
196 & $14.4\pm 1.4$ & & 215 & $8.5\pm 1.1$ & & 165 & $6.7\pm 1.1$ \nl
227 & $12.3\pm 1.2$ & & 245 & $6.8\pm 0.9$ & & 196 & $5.3\pm 0.9$ \nl
257 & $11.4\pm 1.1$ & & 277 & $6.2\pm 0.8$ & & 228 & $4.4\pm 0.7$ \nl
289 & $ 8.4\pm 0.8$ & & 308 & $5.6\pm 0.7$ & & 260 & $3.9\pm 0.7$ \nl
324 & $ 7.7\pm 0.8$ & & 338 & $5.2\pm 0.7$ & & 296 & $2.8\pm 0.5$ \nl
357 & $ 7.7\pm 0.8$ & & 369 & $4.7\pm 0.6$ & & 334 & $2.6\pm 0.4$ \nl
391 & $ 6.1\pm 0.6$ & & 404 & $3.4\pm 0.4$ & & 371 & $2.3\pm 0.4$ \nl
430 & $ 4.6\pm 0.5$ & & 440 & $1.9\pm 0.2$ & & 414 & $1.7\pm 0.3$ \nl
\enddata
\tablenotetext{}{The surface densities are given in units of number per arcmin$^2$ for the sample with $T_1<23.0$ mag.}
\end{deluxetable}

\begin{deluxetable}{cccccr@{$\pm$}lr@{$\pm$}lr@{$\pm$}l}
\tablecolumns{11}
\tablewidth{0pc}
\tablenum{5}
\tablecaption{Structural parameters of the globular cluster system in NGC 4472.}
\tablehead{
\colhead{R [arcsec]} & \multicolumn{3}{c}{Ellipticity} &
\colhead{} & \multicolumn{6}{c}{Position Angle} \\
\cline{2-4}  \cline{6-11} \\
\colhead{}        & \colhead{\sc ALL}           &
\colhead{\sc BGC}           & \colhead{\sc RGC}           &
\colhead{}                  & \multicolumn{2}{c}{\sc ALL} &
\multicolumn{2}{c}{\sc BGC} & \multicolumn{2}{c}{\sc RGC} }

\startdata
 72 & $0.03\pm 0.04$ & $0.05\pm 0.05$ & $0.07\pm 0.04$ & & 135 & 36 & 184 & 29 & 113 & 15 \nl
120 & $0.12\pm 0.06$ & $0.05\pm 0.08$ & $0.06\pm 0.07$ & & 134 & 14 & 106 & 47 & 143 & 34 \nl
168 & $0.11\pm 0.07$ & $0.26\pm 0.08$ & $0.06\pm 0.06$ & & 144 & 18 & 124 & 10 & 151 & 32 \nl
216 & $0.17\pm 0.07$ & $0.28\pm 0.08$ & $0.12\pm 0.09$ & & 153 & 14 & 132 & 10 & 156 & 23 \nl
264 & $0.20\pm 0.08$ & $0.08\pm 0.09$ & $0.21\pm 0.09$ & & 149 & 12 & 137 & 35 & 143 & 14 \nl
312 & $0.14\pm 0.07$ & $0.03\pm 0.10$ & $0.24\pm 0.09$ & & 132 & 15 &  20 & 83 & 135 & 12 \nl
360 & $0.19\pm 0.06$ & $0.13\pm 0.07$ & $0.28\pm 0.09$ & & 149 & 10 & 165 & 16 & 137 & 11 \nl
408 & $0.20\pm 0.04$ & $0.08\pm 0.09$ & $0.31\pm 0.07$ & & 151 &  6 & 184 & 34 & 141 & 8  \nl
456 & $0.13\pm 0.08$ & $0.05\pm 0.11$ & $0.31\pm 0.06$ & & 164 & 21 & 205 & 62 & 144 & 6  \nl
\enddata
\end{deluxetable}

\clearpage

\begin{figure}[1]
\plotone{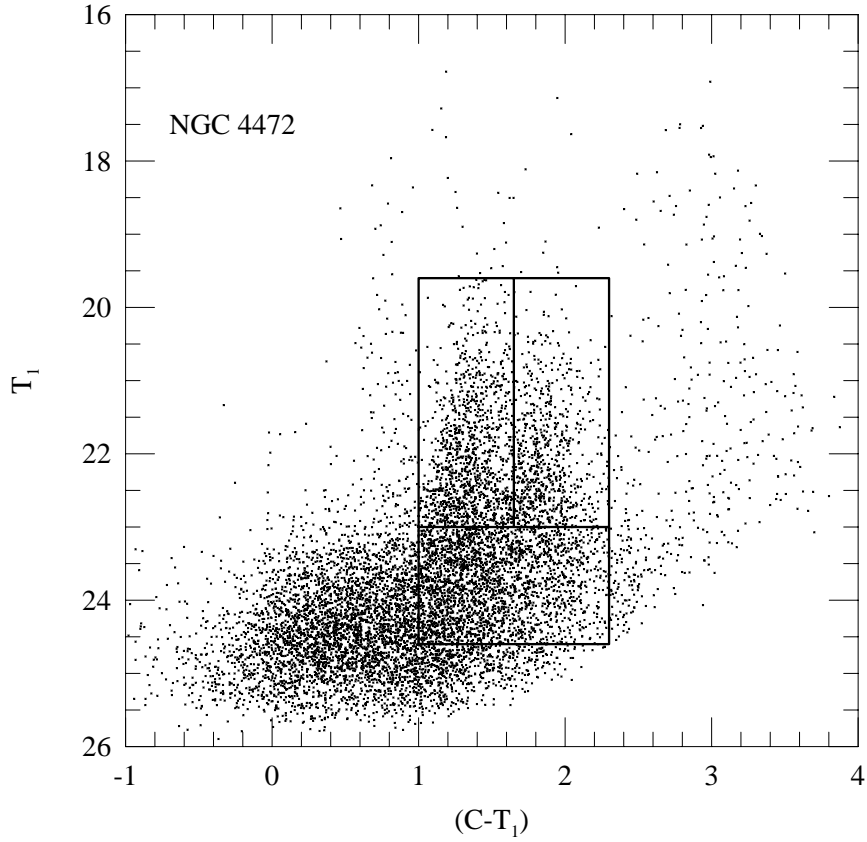}
\vspace{-7truecm}
\figcaption{
$T_1$ vs. $(C-T_1)$ diagram of all objects from Paper I.
The boundaries for the blue globular clusters (BGCs) and red globular clusters
(RGCs), as defined in Paper I,
are marked by the boxes. The middle horizontal line marks the lower
limit of our bright
GC sample used to study the GCS spatial structure, while the lower line indicates
the completeness limit of our photometry.
}
\end{figure}

\begin{figure}[2]
\vspace{-2truecm}
\plotone{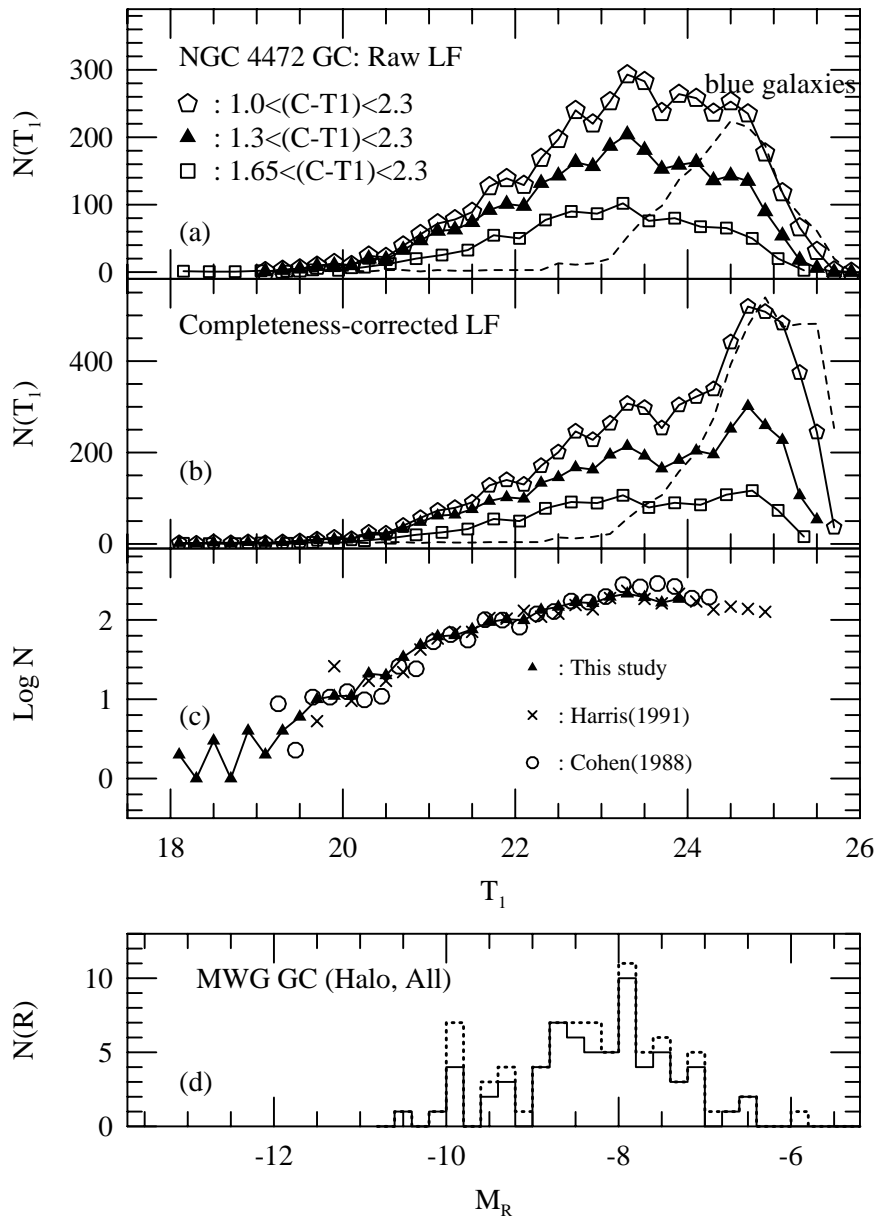}
\vspace{-1truecm}
\figcaption{ (a) $T_1$ luminosity functions of the globular clusters in NGC 4472 before completeness correction
(the pentagons: $1.0<(C-T_1)<2.3$; the triangles: $1.3<(C-T_1)<2.3$;and
the squares: $1.65<(C-T_1)<2.3$).
The dashed line represents the luminosity function of the
blue objects with $0.2<(C-T_1)<0.8$ which are mostly background galaxies.
(b) Completeness-corrected luminosity functions. Note that our results have
not been corrected for background contamination, and are unreliable for $T_1>
23.8$;
(c) Comparison of our results (the triangles) with previous studies.
The circles and crosses represent the $g$-band data given by Cohen (1988) and
the $B$-band data given by Harris \etal (1991), respectively.
These two sets of data were arbitrarily moved along the vertical direction
to match with ours around $T_1=23$ mag, 
and along the horizontal direction
according to the magnitude relations as described in the text. We have limited
our data to $T_1<23.8$.
(d) $R$ luminosity function of the Galactic globular clusters
(the solid line: halo globular clusters,
and the dotted line: both halo and disk globular clusters).}
\end{figure}

\begin{figure}[3]
\vspace{3truecm}
\plotone{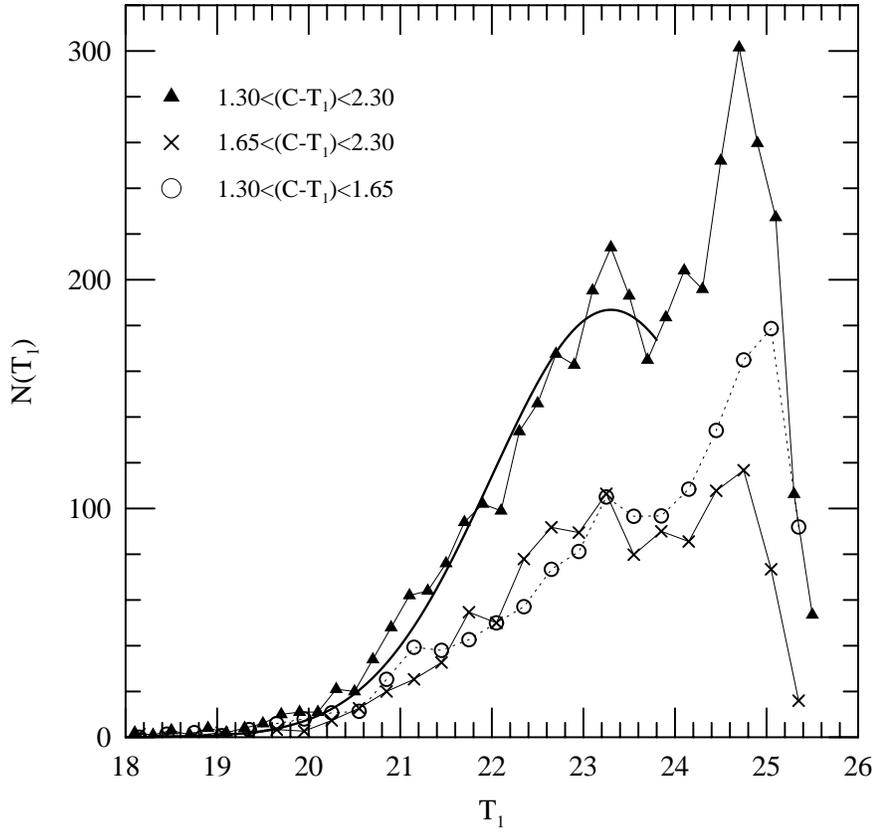}
\vspace{-7truecm}
\figcaption{Comparison of the luminosity functions for
the GCs with $1.3<(C-T_1)<2.3$ (triangles), the BGCs with $1.3<(C-T_1)<1.65$
 (the circles)
 and the RGCs (the crosses). The solid line represents a Gaussian fit to the luminosity
function of the globular clusters with $1.3 <(C-T_1)<2.3$, and  has a peak
at $T_1=23.31$ mag and $\sigma=1.3$.}
\end{figure}

\begin{figure}[4]
\vspace{3truecm}
\plotone{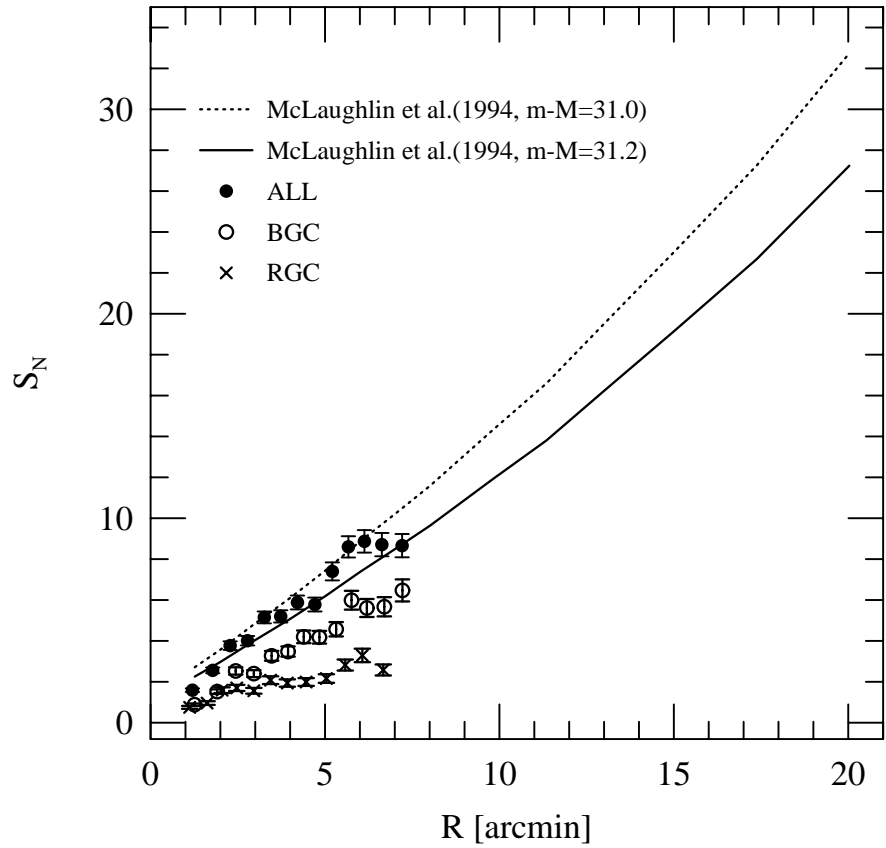}
\vspace{-7truecm}
\figcaption{Radial variation of the local specific frequency.
The dashed line represents the results given by McLaughlin \etal (1994),
and the solid line represents the same results adjusted for the
distance modulus of NGC 4472 obtained in this study. Open   circles are for
the BGCs, crosses for RGCs and filled circles for the combined sample. Error
bars include both Poisson errors in the GC counts and errors in the surface
photometry.}
\end{figure}

\begin{figure}[5]
\vspace{0truecm}
\plotone{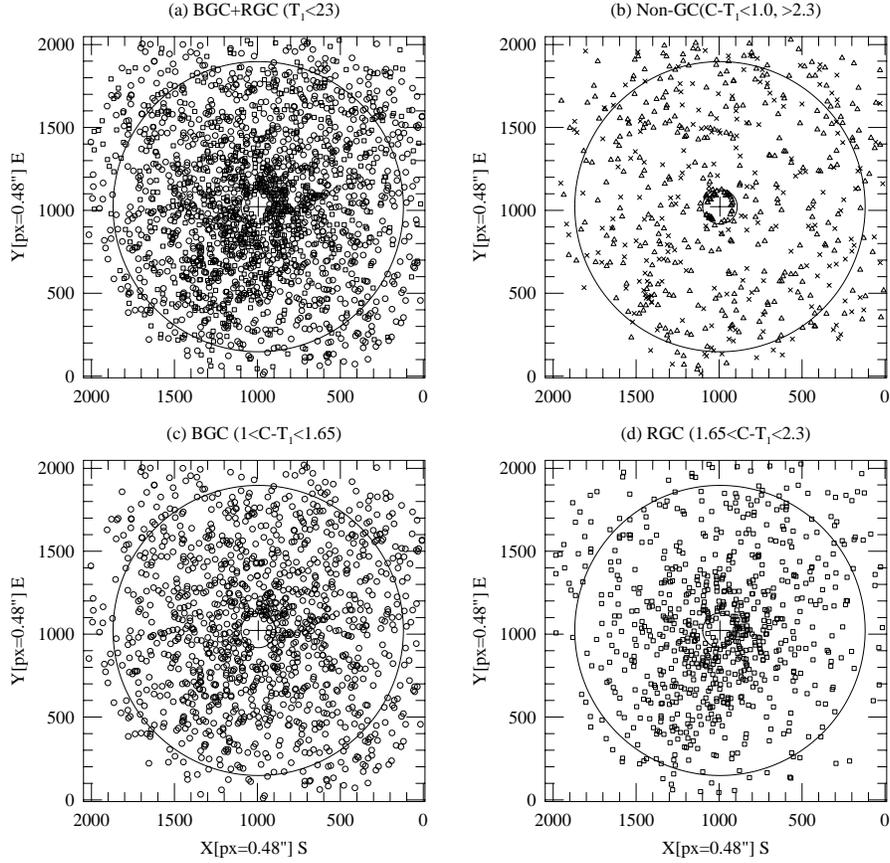}
\vspace{-5truecm}
\figcaption{Positions of the measured objects with $T_1<23$ mag in the field.
The cross represents the center of NGC 4472, and the circles with radii
of 63 pixels (=$50''$) and 875 pixels (=$420''$) represent the boundary
of the region used for the final sample in this study.
(a) All globular clusters. (b) Very blue (the crosses) or very red (the triangles) objects
which are foreground stars or background galaxies. (c) The BGCs. (d) The RGCs.}
\end{figure}

\begin{figure}[6]
\plotone{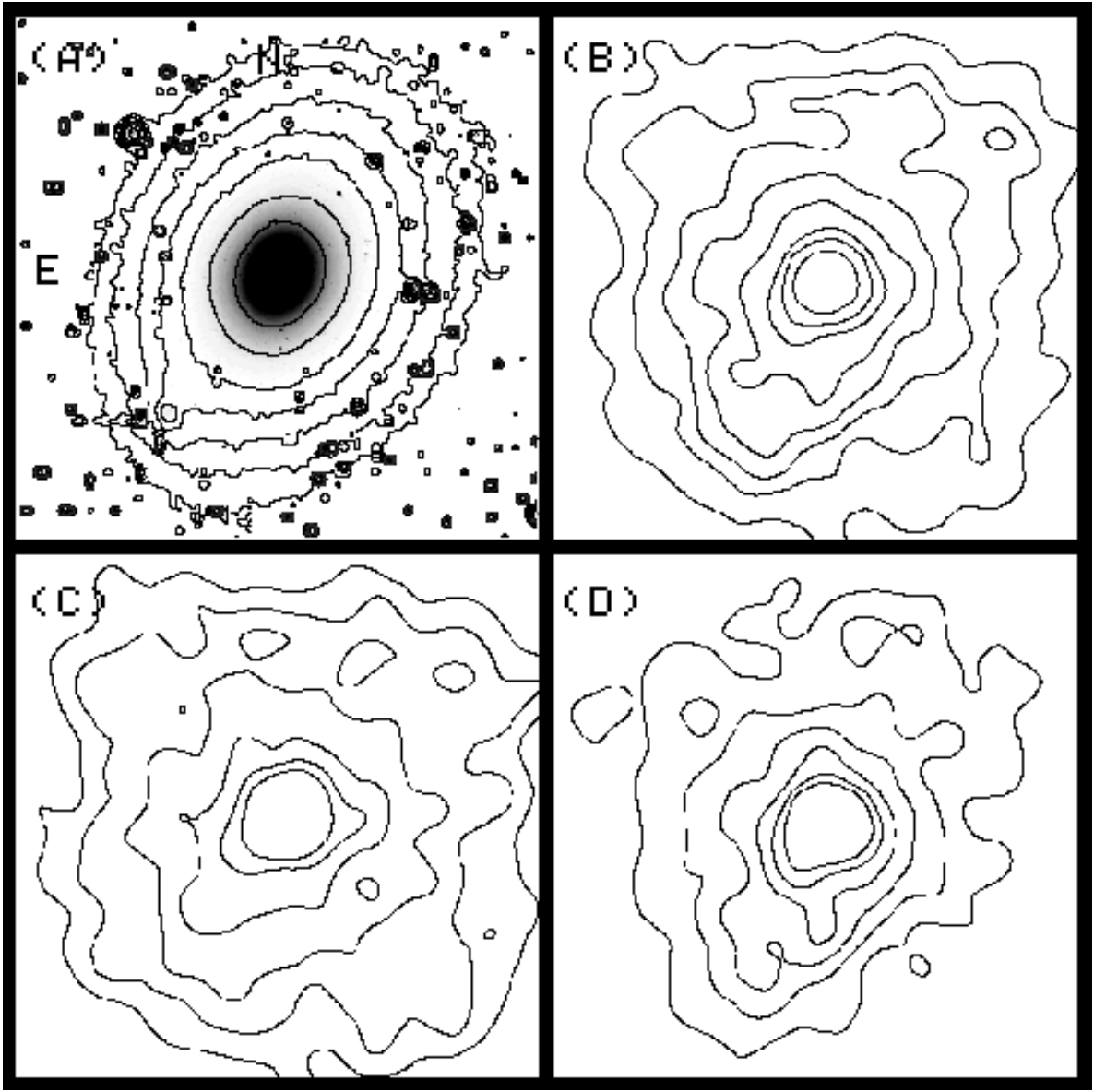}
\vspace{1truecm}
\figcaption{A  greyscale map of a short exposure $T_1$ CCD image of
NGC 4472 (a),
and surface number density maps for the globular clusters with $T_1<23$ mag
((b) the entire GCs, (c) the BGCs and (d) the RGCs).
The contour levels are  3.1, 6.3, 8.8, 12.5, 17.5, 25.0, 31.3, and 43.8
objects/arcmin$^2$ for (b) and
1.9, 3.8, 6.3, 9.4, 12.5 and 15.6 objects/arcmin$^2$
for (c) and (d).
}
\end{figure}

\begin{figure}[7]
\vspace{2truecm}
\plotone{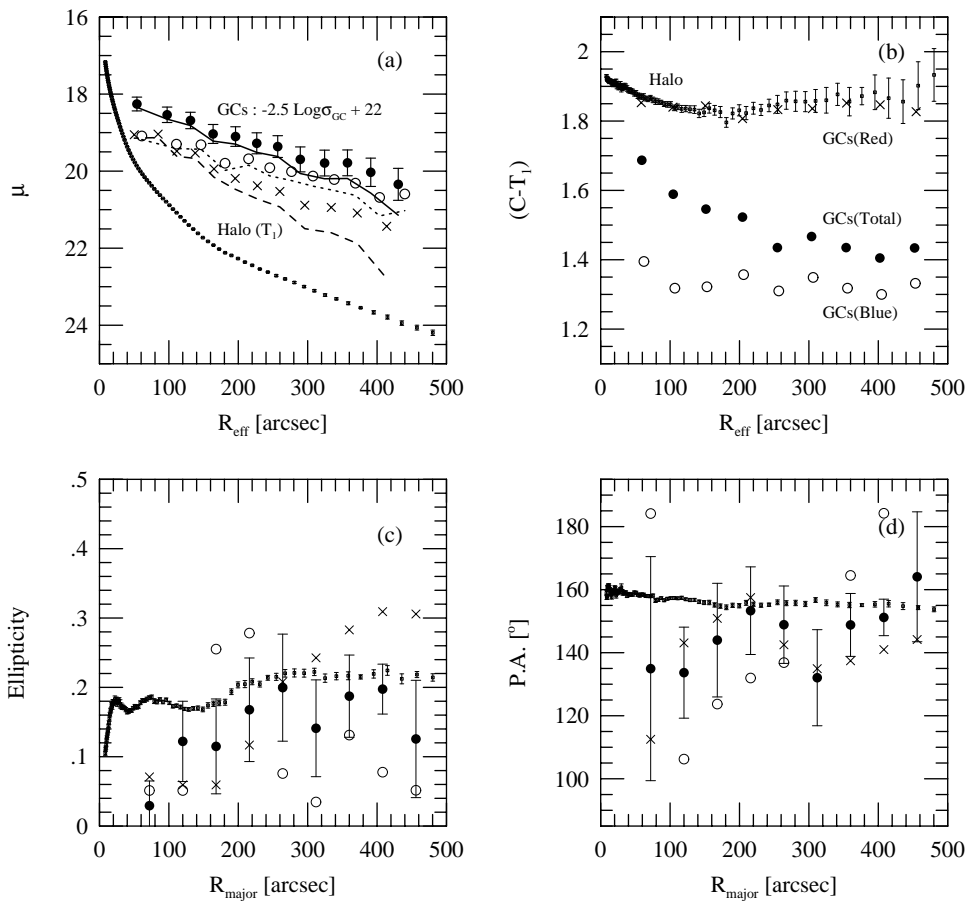}
\vspace{-6truecm}
\figcaption{Radial variations of the surface number density, color, ellipticity,
and position angle of the globular clusters in NGC 4472.
The filled circles, open circles and crosses represent, respectively,
the entire GCS, the BGCs and the RGCs. Also shown as small squares with error bars
are the surface brightness and other parameters for the halo light of NGC 4472.
In (a), both the raw data points as
well as the background-subtracted data (solid, dotted and dashed lines,
respectively) are shown.}
\end{figure}

\begin{figure}[8]
\vspace{-2truecm}
\plotone{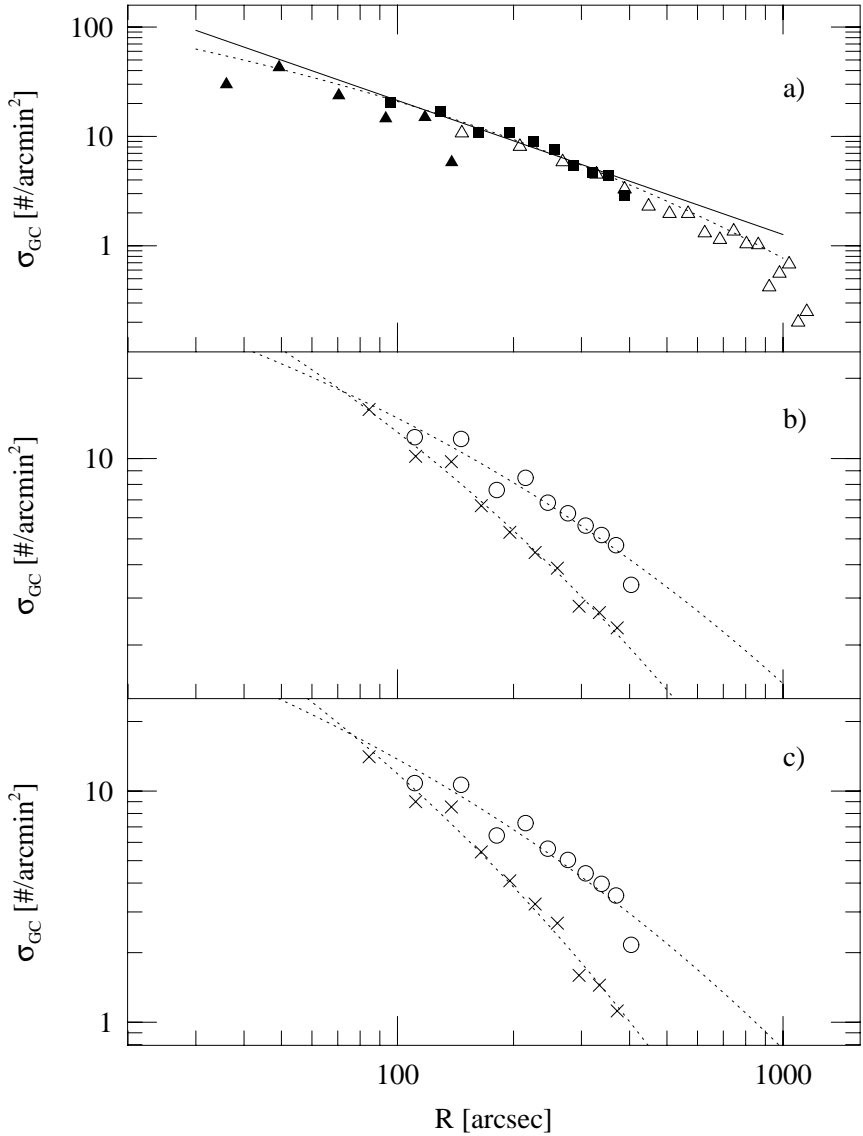}
\vspace{-2truecm}
\figcaption{
(a) Comparison of this study and previous studies
of the radial variation of the surface number density
of the globular clusters in NGC 4472.
(the filled squares: the entire GCS with $T_1 <22.85$ mag  in this study after
background subtraction;
the open triangles: Harris (1986);
the filled triangles: Harris \etal (1991)).
Note that we have shown here only globular clusters brighter than $T_1=22.85$ mag
which is the limiting magnitude of the Harris(1986)'s data, while the limiting
magnitude of the Harris \etal (1991)'s data is $T_1\sim 23.24$ mag.
The solid line and dotted line represent the fits with a power law and
a deVaucouleurs law, respectively.
(b) Comparison of the radial profiles of the metal-poor GCS (the open circles)
 and metal-rich GCS (the crosses) with $T_1<23$ mag, with no background subtraction.
The dotted lines represent the fits with a deVaucouleurs law.
(c) Comparison of the radial profiles of the metal-poor GCS (the open circles)
 and metal-rich GCS (the crosses) with $T_1<23$ mag from which the background
level was roughly subtracted.
The dotted lines represent the fits with a deVaucouleurs law.
}
\end{figure}

\begin{figure}[9]
\vspace{4truecm}
\plotone{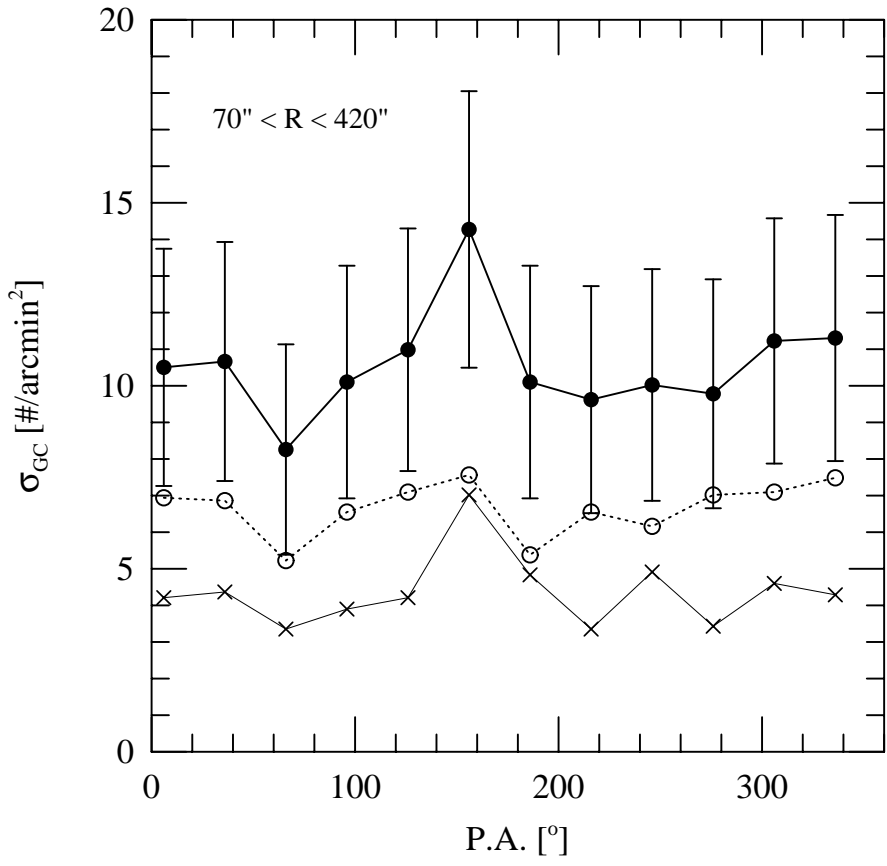}
\vspace{-7truecm}
\figcaption{Azimuthal variations of the surface number density of
the globular clusters with $70''<r<420''$.
The filled circles, open circles and crosses represent, respectively,
the entire GCs, the BGCs and the RGCs.}
\end{figure}

\end{document}